\documentclass[final,1p,times]{elsarticle}

\usepackage{amsmath,amssymb,amsfonts,amscd,amsthm}
\usepackage{graphicx}

\usepackage{color,soul}

\begin{document}

\begin{frontmatter}

\title{Short-range interaction in three-dimensional quantum mechanics}
\author
{Taksu Cheon} 
\ead{taksu.cheon@kochi-tech.ac.jp}
\address
{Laboratory of Physics, Kochi University of Technology,
Tosa Yamada, Kochi 782-8502, Japan}
%

\date{August 30, 2010}

\begin{abstract}
We show that it is possible to define shape-independent three-dimensional
short-range quantum interactions in two parameter form for
non-spherical angular momentum channels through double rescaling of 
potential strength.
Unlike the special case of $l=0$, where the zero-range limit 
of the system is renormalizable, 
the effective ranges diverge for $l \ne 0$ channels, 
and the system becomes trivial at zero-size limit.  
It is also shown that the two-parameter representation with finite 
interaction range is useful in analyzing phase shifts and describing resonances 
with accuracy in non-spherical scatterings.
\end{abstract}


\end{frontmatter}


\section{Introduction}
It is now well known that the quantum point interactions in one dimension 
form a four-parameter family \cite{AG05}.  Its natural extension is a quantum mechanics of
singular vertex of degree $n$ which is described by $n^2$ complex parameters \cite{KS99, CE10}.
It appears to be something of a mystery that quantum point interactions in 
higher dimension than one lack the richness of one dimension, and are strictly
spherical in nature.
The reasoning behind it has been that, for non-spherical $l > 0$ channels,
the existence of centrifugal barrier makes one of two independent free solutions
non-square integrable at the origin, and this leaves only non-interacting boundary 
condition \cite{AG05}.
%
However, this argument does not appear physically convincing 
because all scattering solutions are non-normalizable.
Since we have no way of observing the wave function near the origin, square integrability
is not really an issue if it is properly regularized by a cutoff.   The real issue is whether
we can formulate a disappearing limit of cutoff in such a way that
physical observables are well defined in cutoff-independent manner \cite{JA95}. 

In this note, we consider the short-range limit of three-dimensional quantum 
scattering by small obstacle, and show that it is possible to obtain 
well-defined scattering formula written in terms of
potential strength and effective range.
The absence of zero-range interaction in all channels except $l=0$ is 
expressed as the divergence of effective range in the zero-size limit,
giving a clear physical reasoning.
We also demonstrate, through numerical calculations, that, at finite range, 
the scheme is effective in describing low-energy  non-spherical scatterings.

%
\section{Small size limit of spherical obstacle}

Let us consider a quantum particle scattered off a small object sitting 
at the origin of a three-dimensional space described by the Hamiltonian
\begin{eqnarray}
H = -\frac{{\vec\nabla}^2}{2} + V({\vec r}) .
\end{eqnarray}
We assume that the potential $V({\vec r})$ is zero outside of small region 
$|{\vec r}| \le \Lambda$, and the wave function $\Psi$
satisfies the free Schr{\"o}dinger equation
\begin{eqnarray}
-{\vec\nabla}^2 {\Psi}({\vec r})= k^2 {\Psi}({\vec r}) \quad (|{\vec r}| > \Lambda),
\end{eqnarray}
whose solution we write in partial wave decomposed form
\begin{eqnarray}
{\Psi}({\vec r}) = \sum_{l,m} \frac{\psi_l(r)}{r} Y_{l,m}({\hat r}).
\end{eqnarray}
Each partial wave component $\psi_l$ satisfies the 
Helmholtz equation with centrifugal potential barrier, 
\begin{eqnarray}
-\psi^{\prime\prime}_l(r) + \frac{l(l+1)}{r^2}\psi_l(r) = k^2 \psi_l(r) ,
\end{eqnarray}
whose general solutions can be written as the sum of spherical Bessel
and Neumann functions
\begin{eqnarray}
\psi_l(r) = a_l \left\{ kr \, j_l(kr) \right\} + b_l \left\{ kr \, n_l(kr) \right\} .
\end{eqnarray}
It is often asserted that the Neumann function solution is not permitted
except in the case of $l=0$, since it is too singular at the origin 
and is not square integrable as can be seen from the approximation 
arround $r=0$, which reads
\begin{eqnarray}
\label{absol}
\psi_l(r) 
\approx a_l \, \frac{(kr)^{l+1}}{(2l+1)!!}f_l(kr) - b_l \,  \frac{(2l-1)!!}{(kr)^{l}}g_l(kr)
\qquad (r : {\rm small}) ,
\end{eqnarray}
with
\begin{eqnarray}
&&
f_l(x) \approx 1 - \frac{x^2}{2(2l+3)} +  \frac{x^4}{8(2l+5)(2l+3)}  + \cdots ,
\nonumber \\
&&
g_l(x) \approx 1 + \frac{x^2}{2(2l-1)} +  \frac{x^4}{8(2l-3)(2l-1)}  + \cdots
\qquad (x \to 0).
\end{eqnarray}
However, this argument seems weak, since it would also imply
that we exclude the spherical Bessel function,
because both Bessel and Neumann functions do not fall off at $r \to \infty$;
\begin{eqnarray}
\label{asymptinf}
\psi_l(r) 
\approx a_l  \sin\left({kr-\frac{l\pi}{2}}\right) - b_l   \cos\left({kr-\frac{l\pi}{2}}\right)
\qquad (r \to \infty).
\end{eqnarray}
Here, the issue of square integrability never arises since
the scattering observables are related only to the ratio $a_l/b_l$, and
the inference to $r\to\infty$ is only nominal.
Indeed, we do not measure, for example, such quantity as ratio of probability 
of finding a particle in $r<\Lambda$ and $r>\Lambda$ regions for a given large number $\Lambda$.  Exactly the same argument can be made for {\it small}
$\Lambda$, since we never measure probability amplitude inside a cutoff length
$\Lambda$, and we are only concerned with the efficient description of 
short-ranged potential, in which observables 
do not depend on the cutoff $\Lambda$. 
As long as $\Lambda$ is finite and not exactly zero, 
both terms in (\ref{absol}) has to be kept.
We introduce rescaled amplitudes $\phi_l$ and $\phi^\prime$ defined by
\begin{eqnarray}
\phi^\prime_l = a_l \, \frac{k^{l+1}}{(2l-1)!!} ,
\qquad
\phi_l = - b_l \,  \frac{(2l-1)!!}{k^{l}} .
\end{eqnarray}
We have, at short distance, in low-wave length limit $kr \ll 1$,
\begin{eqnarray}
\psi_l(r) 
\approx \frac{1}{2l+1}\phi^\prime_l  \, r^{\,l+1} f_l(kr)  
 +\  \phi_l  \, \frac{1}{r^{\,l}} g_l(kr)  
\quad (r : {\rm small}) ,
\end{eqnarray}
and
\begin{eqnarray}
\psi_l^\prime(r) 
\approx \frac{l+1}{2l+1}\phi^\prime_l r^{\,l} \left( f_l(kr)+\frac{kr}{l+1}f'(kr) \right) 
  - l\phi_l  \, \frac{1}{r^{\,l+1}} \left( g_l(kr) -\frac{kr}{l}g'(kr) \right) 
\quad (r : {\rm small}) .
\end{eqnarray}
A relation 
\begin{eqnarray}
\frac{a_l}{b_l} =
-  \frac{(2l-1)!!^2}{k^{2l+1}} \frac{\phi^\prime_l}{\phi_l}
\end{eqnarray}
soon becomes handy.  Rewriting the asymptotic form (\ref{asymptinf}), 
we obtain
\begin{eqnarray}
\psi_l(r) 
\approx -\frac{i^{l-1}}{2}(a_l + i b_l)
\left( e^{-ikr}  - (-)^l  \frac{a_l/b_l - i}{a_l/b_l + i} e^{ikr} \right)
\qquad (r \to \infty)
\end{eqnarray}
from which, we can extract the scattering matrix ${\cal S}_l$
defined by $\psi_l(r) \propto e^{-ikr}  - (-)^l  {\cal S}_l e^{ikr}$ in the form
\begin{eqnarray}
{\cal S}_l = e^{2i \delta_l} = \frac{\cot\delta_l+i}{\cot\delta_l-i} 
=  \frac{-a_l/b_l + i}{-a_l/b_l - i} .
\end{eqnarray}
This leads to an expression
\begin{eqnarray}
- k^{2l+1} \cot\delta_l = k^{2l+1} \frac{a_l}{b_l} = -(2l-1)!!^2 \frac{\phi^\prime_l}{\phi_l} .
\end{eqnarray}
We now assume that our obstacle is a sphere of radius $\Lambda$,
and consider the most general boundary condition.
Obviously, the probability current has to conserve at $r=\Lambda$,
and since there is no escaping route in $r<\Lambda$ side, it has to be zero;
\begin{eqnarray}
J_l 
=\frac{i}{2}\left( \psi_l^{\prime *}(r)\psi_l(r)-\psi_l^{\prime}(r)\psi_l^{*}(r) \right)
= 0 .
\end{eqnarray}
This condition is satisfied by wave function with boundary condition
\begin{eqnarray}
\label{robin}
\psi_l^\prime(\Lambda) 
- C_l \psi_l(\Lambda)= 0.
\end{eqnarray}
Here $C_l$ represents the property of the hard surface, $C_l=0$ representing
the Neumann boundary and $C_l=\infty$ the Dirichlet.  
The relation (\ref{robin}) is a generic condition that is
satisfied by any wave function if $C_l$ is allowed to depend on  the incident 
momentum $k$ \cite{BE09}. 
If we consider the scatterings of particles of sufficiently low incident momentum,
the value of the wave function and its derivative at the surface $r =\Lambda$ 
will stay almost the same for any $k$ and can safely be replaced by the ones
for $k=0$.  In other word, for the sace of  
$k \Lambda \lesssim 1$, we can neglect the $k$-dependence, and 
regard $C_l$ as depending only on the cutoff length $\Lambda$.
We focus on whether it is possible to define a set of boundary conditions
that has a ``good'' $\Lambda \to 0$ limit.

Before considering such limit, we ask how we actually 
arrive at the condition (\ref{robin}) 
from finite range potentials.  
There are number of ways, all involving  the rescaling of potential strength.
We take brief looks at two of them.
First, let us consider Dirichlet boundary at $r=0$, namely 
\begin{eqnarray}
\label{odir}
\psi_l(0)=0, 
\end{eqnarray}
and a poential
given by Dirac's $\delta(x-\Lambda)$ with the strength $v$ which might
depend on the distance $\Lambda$ that is represented 
by the connection condition
\begin{eqnarray}
\psi_l^\prime(\Lambda_+) - \psi_l^\prime(\Lambda_-) 
= 2 v(\Lambda) \psi_l(\Lambda) .
\end{eqnarray}
With the obvious relations $\psi_l^\prime(\Lambda_-) \approx \psi_l^\prime(0)$ 
and $\psi_l(\Lambda_-)$ $\approx \psi_l(0) 
+\Lambda \psi_l^\prime(\Lambda_-)$ for small $\Lambda$,
this leads to
\begin{eqnarray}
\label{robinx}
\psi_l^\prime(\Lambda) 
- \left( \frac{1}{\Lambda}+2 v(\Lambda) \right) \psi_l(\Lambda)= 0.
\end{eqnarray}
The choice
\begin{eqnarray}
2 v(\Lambda) = -C_l(\Lambda)-\frac{1}{\Lambda}
\end{eqnarray}
obviously gives the condition (\ref{robin}).
The second way to obtain this same condition also starts from the Dirichlet 
boundary at the origin, (\ref{odir}), and the constant finite potential of 
radius $\Lambda$ given by
\begin{eqnarray}
V(r) = -U(\Lambda) \Theta(\Lambda-r) .
\end{eqnarray}
This gives the wave function at $r=\Lambda$ as 
$\psi(\Lambda) = \psi'(0) \frac{ \sin{\tilde k} \Lambda }{ \tilde k }$
and
$\psi'(\Lambda) = \psi'(0) \cos{\tilde k} \Lambda$  , 
with
\begin{eqnarray}
{\tilde k(\Lambda)} = \sqrt{k^2+2U(\Lambda)} \approx \sqrt{2U(\Lambda)},
\end{eqnarray}
as we let $U$ far larger than $k$ in the small $\Lambda$ limit.
We then have the boundary condition
\begin{eqnarray}
\label{robiny}
\psi_l^\prime 
- \left( {\tilde k} \cot {\tilde k}\Lambda  \right) \psi_l(\Lambda)= 0.
\end{eqnarray}
Now, by choosing ${\tilde k(\Lambda)}$ to satisfy
\begin{eqnarray}
{\tilde k(\Lambda)} \cot \left\{ {\tilde k(\Lambda)}\Lambda \right\} = -C_l(\Lambda), 
\end{eqnarray}
we again obtain the condition (\ref{robin}).

The general sphere surface boundary condition (\ref{robin}) is rewritten 
in terms of $\phi_l$ and $\phi_l^\prime$, as
\begin{eqnarray}
\phi_l^\prime
= -\frac{2l+1}{\Lambda^{2l+1}}
  \left( \frac{l}{\Lambda}-C_l(\Lambda)-k\frac{g'_l(k\Lambda)}{g_l(k\Lambda)} \right)
  \left( -\frac{l+1}{\Lambda}-C_l(\Lambda)-k\frac{f'_l(k\Lambda)}{f_l(k\Lambda)} \right)^{-1}
  \frac{g_l(k\Lambda)}{f_l(k\Lambda)} \phi_l  .
\end{eqnarray}
This condition has non-trivial limit $\phi_l \ne 0$
only when $C_l(\Lambda)$ is rescaled with the $\Lambda$-dependence given by
\begin{eqnarray}
\label{snontriv}
C_l(\Lambda)=\frac{l}{\Lambda} + \chi_l \Lambda^{2l} .
\end{eqnarray}
With
\begin{eqnarray}
&&
\frac{g'_l(x)}{g_l(x)} \approx \frac{x}{2l-1}+\frac{x^3}{(2l-1)^2(2l-3)} \cdots ,
\nonumber \\
&&
\frac{f'_l(x)}{f_l(x)} \approx -\frac{x}{2l+3} -\frac{x^3}{(2l+3)^2(2l+5)}\cdots ,
\nonumber \\
&&
\frac{g_l(x)}{f_l(x)} \approx 1+\frac{(2l+1)x^2}{(2l-1)(2l+3)} 
 +\frac{(l+3)(2l+1)x^4}{(2l-3)(2l+3)^2(2l+5)}\cdots ,
\end{eqnarray}
we have a boundary condition given by
\begin{eqnarray}
\phi_l^\prime  + X_l(k, \Lambda) \phi_l= 0 ,
\end{eqnarray}
in terms of rescaled strength 
\begin{eqnarray}
\label{lexp}
X_l(k, \Lambda) 
=\left( \chi_l +\frac{k}{\Lambda^{2l}}\frac{g'_l(k\Lambda)}{g_l(k\Lambda)} \right)
  \left( 1-\frac{k\Lambda}{2l+1}\frac{f'_l(k\Lambda)}{f_l(k\Lambda)} \right)^{-1}
  \frac{g_l(k\Lambda)}{f_l(k\Lambda)} ,
\end{eqnarray}
whose low energy expansion is 
\begin{eqnarray}
X_l(k, \Lambda)   
\approx \chi_l+  \frac{k^2}{(2l-1)\Lambda^{2l-1}} 
+ \frac{k^4}{(2l-3)(2l-1)\Lambda^{2l-3}}   + \cdots .
\end{eqnarray}
The leading term, $\chi_l$ is {\it cutoff independent}, 
but the next order term {\it diverges
in $\Lambda \to 0$ limit}.
The strict zero-range limit of any three dimensional potential
in all channels other than $l=0$ results in the boundary condition
$\phi_l = 0$ which signifies the absence of interaction $\delta_l = 0$.
It is easy to check that any choice of $C_l(\Lambda)$ other than 
(\ref{snontriv}) leads
the trivial limit $\phi_l = 0$, which is equivalent to  the choice $\chi_l=0$.
%
%
Thus we now have a very physical interpretation of the
well established fact that, in three dimension,  the zero-range force is trivial
except in the spherical channel $l=0$.

Although the strict zero-range limit of $l \ne 0$ quantum interaction gives the
trivial result, this does not preclude the meaningful use of low-energy expansion
(\ref{lexp}) for small but finite $\Lambda$.
Taking the first two terms in the expansion, the scattering matrix is given by
\begin{eqnarray}
\label{gensl}
{\cal S}_l(k) =  -\frac{k^{2l+1}+ \textrm{i} (2l-1)!!^2 
\left(\chi_l + \frac{1}{(2l-1)\Lambda^{2l-1}} k^2\right) }
 {k^{2l+1} - \textrm{i} (2l-1)!!^2 
 \left(\chi_l + \frac{1}{(2l-1)\Lambda^{2l-1}} k^2\right) } ,
\end{eqnarray}
and the scattering phase shift $\delta_l$ is given in the form 
\begin{eqnarray}
\label{sscat}
-\frac{k^{2l+1}}{(2l-1)!!^2}\cot\delta_l =  \chi_l
 + \frac{1}{(2l-1)\Lambda^{2l-1}} k^2 .
\end{eqnarray}
This is a generalization of shape-independent low energy scattering formula
with scattering length and effective range for $s$-wave \cite{NE66} to general value
of angular momentum.
We stress that $\chi_l$ is a quantity obtained by two steps of rescaling procedure 
from a regular finite potential.
%
%

The analytic properties of the scattering matrix ${\cal S}_l(k)$ 
is determined by its poles, which are given as the solution of the equation
\begin{eqnarray}
\label{epole}
 \frac{\textrm{i}}{(2l-1)!!^2} k_{pole}^{2l+1} + 
 \frac{1}{(2l-1)\Lambda^{2l-1}} k_{pole}^2 +  \chi_l = 0 .
\end{eqnarray}
When a pole is located on the positive side of the imaginary axis, it represents
a bound state, and if it is located near the real axis with small negative imaginary 
component, it represents the scattering resonance.  Such resonances can arise  
for $l>0$, due to the ``trapped'' state by centrifugal barrier.
%
%
When $\chi_l$ is zero, a pole exits at $k=0$, representing the zero-energy bound state.
When,on the other hand, the quantity $\Lambda^{2l-1}$ is very large, 
the second term in (\ref{epole}) can be
dropped, and an explicit formula for poles is obtained in the form
\begin{eqnarray}
k_{pole} =  
(2l-1)!!^\frac{2}{2l+1}{\chi}^{\frac{1}{2l+1}}e^{i\pi\frac{4p+1}{4l+2}}
\quad(p=1,...,2l+1) .
\end{eqnarray}
A bound state is represented by a pole on the positive imaginary axis.
If the coupling $\chi$ is positive, $\chi>0$, a bound state is found 
only for even $l$ with $p=\frac{l}{2}$.
If, one the other hand, we have for $\chi<0$, 
a bound state exist only for odd $l$ with the choice $p=\frac{l-1}{2}$. 

For $l>0$, there is a possibility of having a pole near the 
(positive imaginary side of) real axis, which represent the 
{\it resonant scattering} due to the ``trapped'' state by centrifugal barrier.
This occurs for and $\chi >0$, and the resonance pole position is given by
\begin{eqnarray}
k_{res.pole} =  
(2l-1)!!^\frac{2}{2l+1} {\chi}  ^{\frac{1}{2l+1}} 
e^{\frac{\pi \textrm{i}}{4l+2} }.
\end{eqnarray}
The resonance occurs  on the real axis at the value
\begin{eqnarray}
k_{res} =  
(2l-1)!!^\frac{2}{2l+1} {\chi}  ^{\frac{1}{2l+1}} 
\cos{\frac{\pi}{4l+2} }.
\end{eqnarray}

\section{Numerical examples}

In this section, we consider the quantum scattering by short-range interactions in $l\ne 0$
by numerical means, and show the usefulness of our shape-independent scattering
formula.
We examine the example of $p$-wave scattering with the choice $l=1$.
The range parameter is set to $\Lambda=0.1$.
We look at three examples of $C_1=9.75$, $C_1=9.999$, and
$C_1=10.25$ as the boundary parameter $C_1$ in (\ref{robin}).
These correspond to $X_1=-25$, $X_1=-0.1$,
and $X_1=25$, each amounting to the weekly attractive, strongly attractive,
and weekly repulsive potentials, respectively.
%
%
\begin{figure}[h]
\center{
\includegraphics[width=4.2cm]{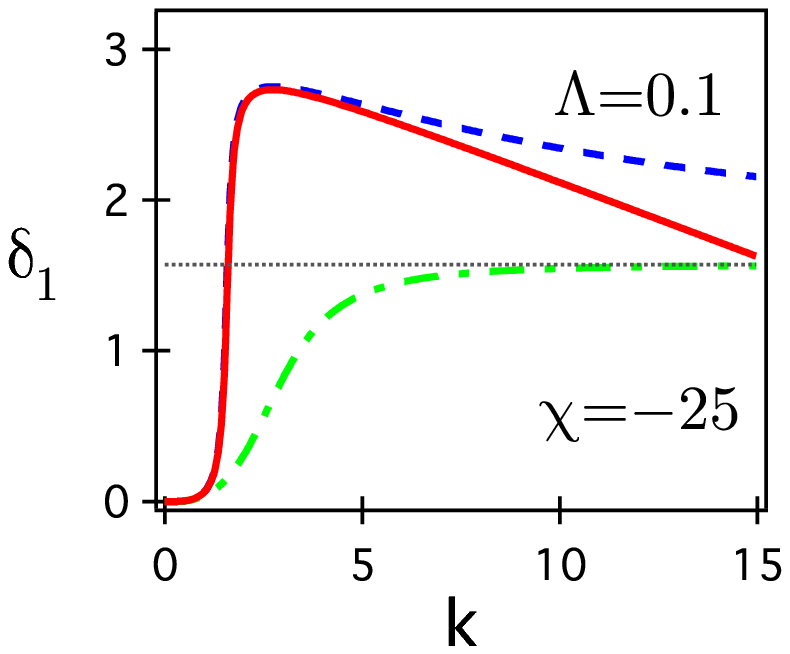}
\includegraphics[width=4.2cm]{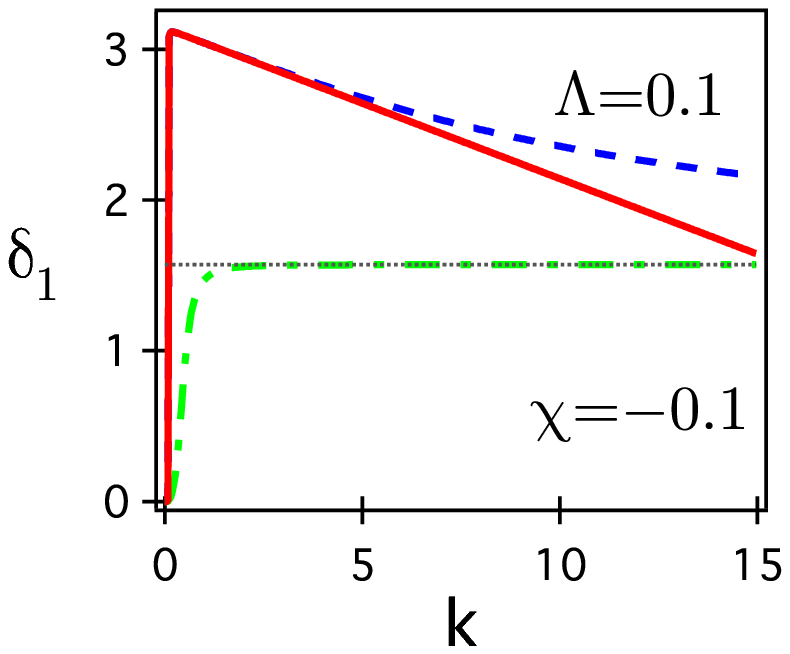}
\includegraphics[width=4.2cm]{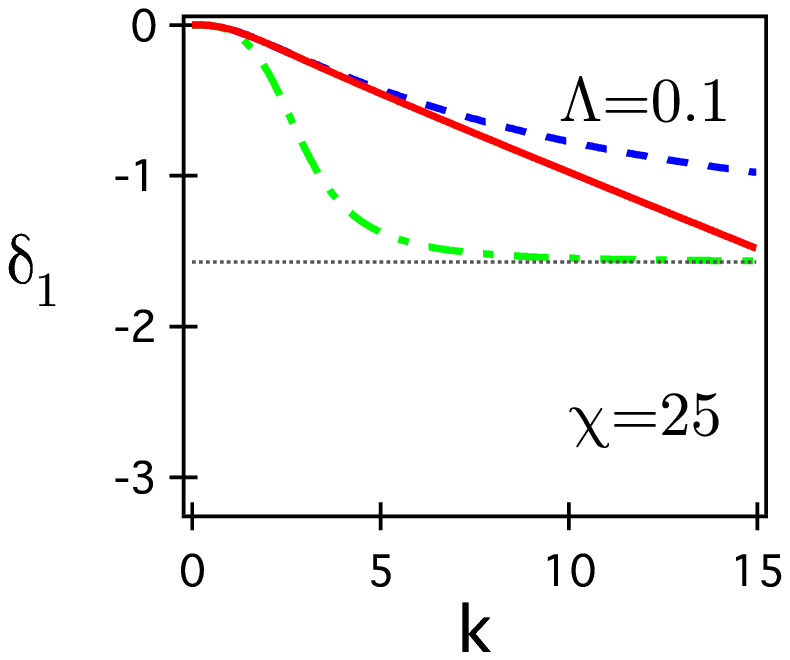}
}
\label{fig1}
\caption
{
Scattering phase shift for $l=1$ short range potential $\delta_1$ with
generalized boundary condition given by $\Lambda$ and $\chi$.  The
solid lines represent the full results, while the dashed line the short range 
approximation with renormalized strength and generalized effective range.
The calculations without the generalized effective range is represented
by dashed-dotted line.
The resonance found in $\chi_1=-25$ and $\chi_1=-0.1$ are related to the
poles of the scattering matrix, which, according to
(\ref{epole}), are given by $k_{pole}=1.5-0.12 \textrm{i}$ and 
$k_{pole}=0.10-0.0005 \textrm{i}$, respectively.
}
\end{figure}
In figure 1, the scattering phase shifts $\delta_1$ is plotted as a function of
the incident momentum $k$.
The results of the full solution with direct calculation from the boundary condition (\ref{robin}) 
are shown in full lines, while the results of the two-parameter shape-independent
scattering formula (\ref{sscat}) are shown with dashed lines.
They are in excellent agreement up to $k\Lambda \approx 0.5$ as expected in the 
preceding analysis.  
The important point to note is the accuracy with which the resonance
is described.
The dashed-dotted lines are the results of ``wrong'' zero-range forrmula
using only the first term of (\ref{sscat}), which is plotted just to show the critical importance
of the $k^2$ term.  
Calculations with higher $l$ than one also yield very similar results showing the
power of shape-independent scattering formula  (\ref{sscat}). 
Our formula should be useful in the analysis of nuclear and atomic scattering experiments.

\section{Conclusion}

It has been established for long time that the quantum zero-range interaction in 
dimension three is trivial except in the spherical channel $l=0$.  
However, the convincing physical reasoning is given in this work for the first time.
In the process of the analysis, we have found a simple general formula for
the short-range scattering for any angular momentum $l$.
Through the numerics, we have established 
that this formula indeed is a meaningful generalization of 
well known result of $l=0$, and that is in spite of the fact that it looses 
the meaning entirely in strict $\Lambda \to 0$ limit because of the divergence. 

The triviality of three-dimensional zero-range interaction is 
in stark contrast against the richness of one-dimensional zero-range
interaction, for which there are infinite variety described by four-parameter family.
A similar analysis to this work for the two-dimensional short-range interaction
is to be done, if any necessity arises for the examination of
quantum two-dimensional scattering.
  
\section*{Acknowledgments}

We would like to thank Dr. O. Turek for useful discussions.
The research has been supported  by the 
Japanese Ministry of Education, Culture, Sports, Science and Technology 
under the Grant number 21540402.


\end{document}